\documentclass[11pt,letter]{article}
\usepackage[top=1in,bottom=1in,left=1in,right=1in]{geometry}
\pdfoutput = 1
\usepackage[noEucal]{main}
\usepackage{aas_macros}
\usepackage[utf8]{inputenc}
\makeatletter \g@addto@macro\@floatboxreset\centering \makeatother
\usepackage{graphicx}
\usepackage[scr=boondoxo]{mathalfa}
\usepackage[bbgreekl]{mathbbol}
\DeclareSymbolFontAlphabet{\mathbbm}{bbold}
\usepackage[font={small,it}]{caption}

\def\CPT{\msR}
\def\SO{{\mathsf S}\msO}
\def\SCO{{\mathscr O}}
\def\ISO{\msI{\mathsf S}\msO}

\def\id{{\mathbbm 1}}

\def\V{{\vec{V}}}

\begin{document}
\begin{titlepage}
\unitlength = 1mm
\ \\
\vskip 3cm
\begin{center}

{\LARGE{\textsc{Celestial Conformal Primaries in Effective Field Theories}}}
\vspace{0.8cm}

Prahar Mitra

\vspace{1cm}
{\it Institute for Theoretical Physics, University of Amsterdam,
Science Park 904, Postbus 94485, 1090 GL Amsterdam, The Netherlands}

\vspace{0.8cm}

\begin{abstract}
Scattering amplitudes in $d+2$ dimensions can be recast as correlators of conformal primary operators in a putative holographic CFT$_d$ by working in a basis of boost eigenstates instead of momentum eigenstates. It has been shown previously that conformal primary operators with $\Delta \in \frac{d}{2} + i {\mathbb R}$ form a basis for massless one-particle representations. In this paper, we consider more general conformal primary operators with $\Delta \in {\mathbb C}$ and show that completeness, normalizability, and consistency with CPT implies that we must restrict the scaling dimensions to either $\Delta \in \frac{d}{2} + i {\mathbb R}$ or $\Delta \in {\mathbb R}$. Unlike those with $\Delta \in \frac{d}{2} + i {\mathbb R}$, the conformal primaries with $\Delta \in {\mathbb R}$ can be constructed without knowledge of the UV and can therefore be defined in effective field theories. With additional analyticity assumptions, we can restrict $\Delta \in 2 - \mzz_{\geq0}$ or $\Delta \in \frac{1}{2}-\mzz_{\geq0}$ for bosonic or fermionic operators, respectively.
\end{abstract}

\vspace{1.0cm}
\end{center}
\end{titlepage}
\pagestyle{empty}
\pagestyle{plain}
\pagenumbering{arabic}

\tableofcontents

\section{Introduction}

Following the seminal work by Strominger \cite{Strominger:2013jfa}, there has been a renewed interest in quantum gravity in asymptotically flat spacetimes, and in the past decade, significant progress has been made on this front (see \cite{Strominger:2017zoo} and more recently \cite{Pasterski:2021rjz, Raclariu:2021zjz} for an extensive review and references). Much of this development has focussed on understanding the holographic nature of scattering amplitudes in asymptotically flat spacetimes. Early results studied the universal factorization of amplitudes in the soft (zero energy) limit and reinterpreted them as consequences of asymptotic symmetries of asymptotically flat spacetimes. For example, in \cite{He:2014laa, Kapec:2014opa}, the BMS supertranslation \cite{Bondi:1962px, Sachs:1962wk} and superrotation symmetries \cite{Barnich:2011ct} were shown to be equivalent to the leading and subleading soft-graviton theorems \cite{Weinberg:1965nx, Cachazo:2014fwa} respectively and in \cite{He:2014cra, He:2015zea, Campiglia:2015qka, Kapec:2015ena, Campiglia:2019wxe, He:2020ifr}, large gauge symmetries in abelian and nonabelian gauge theories were shown to be equivalent to the leading soft-photon and soft-gluon theorems \cite{Weinberg:1965nx, BERENDS1989595}, respectively. Since then, many new asymptotic symmetries and their corresponding soft theorems have been discovered \cite{Lysov:2014csa, Strominger:2015bla, Dumitrescu:2015fej, Campiglia:2016jdj, Campiglia:2016hvg, Campiglia:2016efb, Laddha:2017ygw, Laddha:2017vfh, Himwich:2019dug, Freidel:2021dfs, Strominger:2021mtt, Ball:2021tmb, Freidel:2021ytz}. Despite these developments, a complete understanding of all the relevant asymptotic symmetries of the problem is still lacking.

Asymptotic symmetries have a natural action on the celestial sphere (the co-dimension two sphere at infinity), but their action on the usual asymptotic states (i.e., momentum eigenstates) is rather non-trivial. By working in a new basis, these symmetries can likely be made manifest and provide us with a more rigorous treatment of the same. One such basis was introduced by Pasterski and Shao \cite{Pasterski:2017kqt}. In $d+2$ spacetime dimensions, these one-particle basis operators are labeled by a parameter $\D\in\mcc$, a point $\vec{x}$ on the celestial sphere $\mss^d$ and a spin index $s$ that transforms in a finite-dimensional irreducible representation of $\SO(d)$.\footnote{We work in the language of creation and annihilation operators instead of states or wavefunctions.} The \emph{conformal primary operators} $\SCO_s(\D,\vec{x})$ make the Lorentz symmetry of the scattering amplitude manifest. More precisely, the authors exploited the isomorphism between the Lorentz group in $\mrr^{1, d+1}$ and the conformal group in $\mrr^d$ and constructed $\SCO_s(\D,\vec{x})$ as the highest-weight operators of the conformal group. The conformal primaries are simply a different basis on the space of asymptotic operators, and so, must be related to the usual creation and annihilation operators $\CO_\s(p)$, which are labeled by a momentum $p^\mu$ satisfying $p^2 = - m^2$ and the spin index $\s$ of a finite-dimensional irreducible representation of the little group. The precise form of this relationship is rather simple for massless operators and is given by a Mellin transform\footnote{In the massless case, the little group is $\SO(d)$, and the index $\s$ can be identified with $s$.}
\begin{equation}
\begin{split}\label{map1}
\SCO_s(\D,\vec{x}) = \int_{\mrr_+} \dt \o \o^{\D-1} \CO_s ( \o {\hat q} ( \vec{x} ) )  , \qquad {\hat q}^\mu(\vec{x}) = \left( \frac{1+|\vec{x}|^2}{2} , \vec{x} , \frac{1-|\vec{x}|^2}{2} \right) . 
\end{split}
\end{equation}
The basis transformation can be inverted by use of the Mellin inversion theorem, which, along with the normalizability w.r.t. the Klein-Gordon norm, requires the restriction
\begin{equation}
\begin{split}
\label{Delta1}
\D \in \CC_P \cong \frac{d}{2} + i \mrr . 
\end{split}
\end{equation}
Scattering amplitudes evaluated in the conformal primary basis are known as celestial amplitudes. In recent years, the structure and properties of celestial amplitudes have been extensively studied \cite{Pasterski:2016qvg, Donnay:2018neh, Pate:2019mfs, Puhm:2019zbl, Pate:2019lpp, Albayrak:2020saa, Gonzalez:2020tpi, Arkani-Hamed:2020gyp, Pasterski:2021dqe, Pano:2021ewd, Bagchi:2022emh, Casali:2022fro, Jorge-Diaz:2022dmy, deGioia:2023cbd}. Despite these works, a few complications with the map \eqref{map1} have yet to be addressed fully.

The first problem is that \eqref{map1} involves an integral over all energies, so the conformal primary operators are highly sensitive to UV physics \cite{Arkani-Hamed:2020gyp}. Not only do we need to have complete knowledge of the UV of the theory, but the high energy behavior of the creation and annihilation operators must be sufficiently soft so that the Mellin integral in \eqref{map1} converges. For instance, a $2\to2$ amplitude involving the exchange of a spin $j$ particle grows as $\o^{2(j-1)}$ at high energies, so the Mellin integral is not well-defined.\footnote{In theories of quantum gravity, amplitudes are expected to be exponentially damped at high energies. For example, in string theory, there are infinitely many states with $j \in \mzz_{\geq0}$, and the $2\to2$ amplitude is of the form $\sum_j a_j(\t) \o^{2(j-1)} \sim \exp [ - f(\t) \o^2 ]$ for some functions $a_j$ and $f$ of the scattering angle $\t$ \cite{Gross:1987kza}.} More generally, celestial amplitudes are not constructible in effective field theories, where the physics is only known up to a finite UV scale $\L_{\text{UV}}$.

A second problem with \eqref{map1} is that the translation generators $P_\mu$ do not have a well-defined action on the conformal primary operators. $P_\mu$ acts on the creation and annihilation operators by adding a multiplicative factor of $\o$ implying that it shifts the scaling dimension of the conformal primary by 1, i.e. $\D \mapsto \D + 1$. Consistency with translational symmetry suggests that for every conformal primary with dimension $\D$, we must allow infinitely many primaries with dimension in $\D + \mzz$. This is, however, in contradiction with the requirement from normalizability \eqref{Delta1}, which forced the scaling dimension to lie on the principal series axis. This contradiction is related to the fact that the integral in \eqref{map1} is over all energies. States with arbitrarily high energy contribute to the conformal primary, so $P_\mu$ is not bounded.\footnote{This is not that big a problem and can be fixed by considering the bounded operator $\exp ( - i a^\mu P_\mu )$ instead.}

A third problem is that many operators of interest in celestial holography do not live on the principal series axis. Rather, they have integer scaling dimensions. For example, the conformally soft operators constructed in \cite{Donnay:2018neh, Pate:2019mfs, Puhm:2019zbl} have $\D=1,0$ and the infinitely many $\msw_{1+\infty}$ generators have $\D \in 2 - \mzz_{\geq 0}$ \cite{Strominger:2021lvk}. As shown in \cite{He:2015zea, Kapec:2016jld, Kapec:2017gsg, Nande:2017dba}, the shadow transform\footnote{This is an integral transform which maps a primary of weight $\D$ to a primary of weight $d-\D$.} of the conformally soft operators gives the stress tensor and conserved current operators, respectively. These conformal primaries with $\D \in 2-\mzz_{\geq0}$ are constructed by starting with the primaries given by \eqref{map1} and then analytically continuing $\D$ to the complex plane. The analytic continuation has poles precisely at $\D \in 2-\mzz_{\geq0}$, and the operators are extracted as the residues at these poles.

In this paper, we address these issues by proposing a modified definition for the conformal primary operators given by
\begin{equation}
\begin{split}\label{map2}
\SCO_s(\D,\vec{x}) = \int_{\CC_\o} \frac{\dt \o}{2\pi i}  \o^{\D-1} \CO_s ( \o {\hat q} ( \vec{x} ) )   . 
\end{split}
\end{equation}
This is quite similar to the one proposed by Pasterski-Shao \eqref{map1}, except that the integration is now over a contour $\CC_\o$ in complex $\o$ space.\footnote{Importantly, we do not assume that $\CC_\o$ is a deformation of the original integration contour used in \eqref{map1}.} To constrain $\CC_\omega$, we impose the following conditions:
\begin{itemize}
\item $\SCO_s(\D,\vec{x})$ is a complete basis.
\item $\SCO_s(\D,\vec{x})$ is consistent with Lorentz and CPT symmetries.
\item $\SCO_s(\D,\vec{x})$ is normalizable.
\end{itemize}
These requirements are, of course, satisfied by the Mellin contour $\mrr_+$ used in \eqref{map1}, but they are also satisfied by the contour $\CC_\L \cong \L e^{i\mrr}$. Furthermore, while these requirements fix $\D \in \CC_P$ if $\o \in \mrr_+$, we will show that if $\o \in \L e^{i\mrr}$, then 
\begin{equation}
\begin{split}
\label{Delta2}
\D \in \mrr . 
\end{split}
\end{equation}
This is the central result of our paper.

The conformal primaries defined with the contour $\CC_\L$ address all three of the aforementioned described issues with \eqref{map1}. The integral in \eqref{map2} is over a bounded contour $|\o|=\L$, so we are only required to have knowledge of physics up to the scale $\L$. Consequently, celestial amplitudes involving this operator can be constructed in effective field theories as long as $\L < \L_\text{UV}$. The boundedness of $\CC_\L$ implies that $P_\mu$ is bounded and, therefore, has a well-defined action on the conformal primaries. More precisely, the translation generators still map $\D \mapsto \D + 1$ as before, but now, due to \eqref{Delta2}, this does not contradict normalizability. Finally, the conformally soft operators at $\D=1,0$ and $\msw_{1+\infty}$ operators at $\D=2-\mzz_{\geq0}$ can be extracted directly from these primaries without any analytic continuation.

In defining the conformal primaries via \eqref{map2}, we made no assumptions regarding the analyticity of the creation and annihilation operators in $\o$. The analytic properties of scattering amplitudes -- and hence of the creation and annihilation operators -- have been extensively studied in the literature (see \cite{Mizera:2023tfe} for a modern review), and using those results, one should be able to restrict the properties of the conformal primaries \eqref{map2} further. For example, using the fact that massless creation and annihilation operators admit a Laurent expansion near $\o=0$, we will show that we can restrict the scaling dimensions of the conformal primaries to $\D \in \mzz$. Conformal primary operators of this type have previously appeared in \cite{Freidel:2022skz, Cotler:2023qwh}. In these works, the authors mirror the discussion of Pasterski-Shao and construct conformal primary wavefunctions $\Uppsi_{\D,\vec{x}}(X)$ which are solutions to the wave-equation but transform as conformal primaries under Lorentz transformations. Using a modified definition of the norm (\!\!\!\cite{Freidel:2022skz} uses the $\mll^2$-norm on Schwarz space and \cite{Cotler:2023qwh} uses the RSW norm), they show that the wavefunctions are normalizable if $\D \in \mzz$. They also show that the wavefunctions with $\D \in \mzz$ form a complete basis in their respective spaces (Schwarz space for \cite{Freidel:2022skz}, and space of Wightman functions for \cite{Cotler:2023qwh}). The methods used in those works are quite different, though many conclusions are identical. This paper presents a third construction of these integer conformal primaries, working entirely in momentum space and using the language of creation and annihilation operators instead of spacetime wavefunctions.

The remainder of this paper is organized as follows. In section \ref{sec:symmetry_review}, we present a lightning review of one-particle representations of the Poincar\'e group in $d+2$ dimensions. In section \ref{sec:group_isomorphism}, we review the isomorphism between the Lorentz group in $\mrr^{1,d+1}$ dimensions and the conformal group in $\mrr^d$. In section \ref{sec:amp_correlators}, we introduce helpful notation that manifests the conformal structure of creation and annihilation operators. In section \ref{sec:Pasterski_Shao}, we briefly review the work of Pasterski and Shao \cite{Pasterski:2017kqt}. The central result of our paper is described and proved in section \ref{sec:constraints}. Finally, we comment on future directions in section \ref{sec:summary}.

\section{Preliminaries}
\label{sec:scat_amp}

The LSZ reduction formula recasts a scattering amplitude in $d+2$ spacetime dimensions (the transition amplitude between asymptotic $in$ and $out$ states) as a time-ordered vacuum correlation function of the operators
\begin{equation}
\begin{split}
\label{LSZ}
\CO_s (p) =  - i \ve_s^A (p) \int_{\mrr^{d+2}} \dt^{d+2} X e^{ - i p \cdot X  } \p^2 \varphi_A ( X ) .
\end{split}
\end{equation}
The operator $\CO_s(p)$ -- which we will from now on refer to as the \emph{plane wave operator} -- is labeled by a momentum $p^\mu$ satisfying $p^2 = - m^2$ and a spin index $s$ that transforms under some finite-dimensional unitary irreducible representation of the little group ($\SO(d+1)$ if $m>0$ and $\ISO(d)$ if $m=0$). The spacetime field $\varphi_A(X)$ is labeled by a spin index $A$ that transforms in some representation of the Lorentz group. The polarization tensor $\ve^A_s(p)$ describes the embedding of the plane wave operator in the field. According to \eqref{LSZ}, $\CO_s(p)$ inserts an outgoing particle in the amplitude if $p^0 > 0$, and it inserts an incoming particle if $p^0 < 0$.

In the rest of this paper, we will restrict our attention to massless operators. We also use matrix notation and drop the spin indices $s$ and $A$.

\subsection{Poincar\'e and CPT Symmetry}
\label{sec:symmetry_review}

Relativistic quantum field theories are invariant under proper orthochronous Lorentz and discrete CPT transformations. These are generated by a unitary linear operator $U(\L)$ and an anti-unitary anti-linear operator $\CPT$, respectively.
The action of these operators on the Hilbert space can be determined from basic representation theory (e.g., see Chapter 2 of Weinberg \cite{Weinberg:1995mt}),
\begin{equation}
\begin{split}
\label{field_LT_CPT}
U(\L) \cdot \varphi ( X ) = D (\L^{-1}) \varphi (\L X ) , \qquad \CPT \cdot \varphi( X ) = S^\star \varphi^\dagger (  \CR X ) ,
\end{split}
\end{equation}
where $D(\L)$ is a representation of the Lorentz group, $\CR^\mu{}_\nu = \diag(-1,+1,\cdots,+1,-1)$ and the matrix $S$ satisfies
\begin{equation}
\begin{split}
S D(\L) S^{-1} = D ( \CR \L \CR^{-1} )~~\forall~~ \L \in \SO^\uparrow(1,d+1) , \qquad S^{-1} = S^\dagger = e^{- \pi i F} S , 
\end{split}
\end{equation}
where $F=0$ if $D$ is bosonic and $F=1$ if it is fermionic. Using \eqref{LSZ}, \eqref{field_LT_CPT} and the properties
\begin{equation}
\begin{split}
\ve (\L p) = \SD (W(\L,p)) \ve (p) D(\L^{-1}) , \qquad \ve(\CR p) = C^{-1} \ve(p) S 
\end{split}
\end{equation}
we can determine the Lorentz and CPT transformation properties of the plane wave operators
\begin{align}
\label{Ok_LT}
U(\L) \cdot \CO (p) &=  \SD(W(\L,p)^{-1}) \CO (\L p) , \qquad \CPT \cdot \CO (p) = C^\star \CO^\dagger (\CR p ) .
\end{align}
Here, $\SD(W)$ is the unitary irreducible finite-dimensional representation of the massless little group $\ISO(d)$, and the matrix $C$ satisfies
\begin{equation}
\begin{split}
\label{Ck_prop}
C \SD(W) C^{-1} = \SD(\CR W \CR^{-1} ) ~~ \forall~~ W \in \ISO(d)  , \qquad C^{-1} = C^\dagger = e^{- \pi i F}  C.
\end{split}
\end{equation}
The little group element $W(\L,p)$ is defined by
\begin{equation}
\begin{split}
\label{W_def}
W(\L,p) \equiv L( \L p )^{-1} \L L(p)  , 
\end{split}
\end{equation}
where $L(p)$ is a Lorentz matrix satisfying
\begin{equation}
\begin{split}
\label{Lp_def}
L(p)^\mu{}_\nu r^\nu = p^\mu , \qquad L(p) = \CR^{-1} L( - \CR  p ) \CR , \qquad r^\mu = \frac{1}{2} (1,0,\cdots,0,1).
\end{split}
\end{equation}
Recall that the little group is the subgroup of $\SO^\uparrow(1,d+1)$ consisting of all elements that preserve the reference momentum $r^\mu$, i.e. it consists of all elements $W \in \SO^\uparrow(1,d+1)$ such that $W^\mu{}_\nu r^\nu = r^\mu$. 

In addition to Lorentz and CPT transformations, quantum field theories are also invariant under translations, which are infinitesimally generated by the Hermitian generators $P_\mu$. This acts as
\begin{equation}
\begin{split}
\label{translation_action}
P_\mu \cdot \varphi(X) =  i  \p_\mu \varphi(X) \quad \implies \quad P_\mu \cdot \CO(p) = - p_\mu \CO(p) . 
\end{split}
\end{equation}

\subsection{Euclidean Conformal Group \texorpdfstring{$\cong$}{iso} Lorentz Group}
\label{sec:group_isomorphism}

The Euclidean conformal group is the space of all diffeomorphisms $\vec{x} \to \vec{V}(\vec{x})$ on $\mrr^d$ satisfying
\begin{equation}
\begin{split}
\label{conf_diff_def}
\dt \vec{V}(\vec{x}) \cdot \dt \vec{V}(\vec{x}) = \O_{\vec{V}}^2(\vec{x}) \dt \vec{x} \cdot \dt \vec{x} , \qquad \vec{x} = x^a = (x^1,\cdots,x^d).
\end{split}
\end{equation}
There are five classes of conformal diffeomorphisms -- translations $\vec{{\tilde n}}_{\vec\xi}$, $\SO(d)$ rotations $\vec{m}_R$, dilations $\vec{a}_\l$, special conformal transformations $\vec{n}_{\vec{ a}}$ and inversions $\vec{i}$. These are summarized in Table \ref{table:conformal_group}. A generic element of the conformal group can be uniquely decomposed as $\vec{V} = \vec{{\tilde n}}_{\vec\xi} \circ \vec{m}_R \circ \vec{a}_\l \circ \vec{n}_{\vec{ a}}$. For each $\vec{V}$, we define
\begin{equation}
\begin{split}
\label{Rchi_def}
[R_{\vec{V}}(\vec{x})]^a{}_b \equiv \O_{\vec{V}}^{-1}(\vec{x}) \p_b V^a(\vec{x}) \in \msO(d) , \qquad \vec{\chi}_{\vec{V}}(\vec{x}) \equiv \frac{1}{2} R_{\vec{V}}(\vec{x}) \vec{\n} \ln \O_{\vec{V}}(\vec{x}) . 
\end{split}
\end{equation}
These quantities are also summarized in Table \ref{table:conformal_group}.
\begin{table}[ht!]
\renewcommand{\arraystretch}{1.25}
\begin{center}
\begin{tabular}{|c|c|c|c|c|}
\hline
Name, $\vec{V}$ & $\vec{V}(\vec{x})$ & $\O_\V(\vec{x})$ & $R_\V(\vec{x})$ & $\vec{\chi}_\V(\vec{x})$ \\
\hline \hline 
Translations, $\vec{\wt n}_{\vec a}$ & $\vec{x} + \vec{ a}$ & $1$ & $\id$ & $\vec{0}$ \\
Rotations, $\vec{m}_R$, $R \in \SO(d)$ & $\overrightarrow{R x}$ & $1$ & $R$ & $\vec{0}$ \\
Dilations, $\vec{ a}_\l$, $\l>0$ & $\l \vec{x}$ & $\l$ & $\id$ & $\vec{0}$ \\
SCT, $\vec{n}_{\vec{a}}$ & {$\frac{\vec{x} + |\vec{x}|^2 \vec{ a} }{ 1 + 2 \vec{ a} \cdot \vec{x} + | \vec{ a} |^2 |\vec{x} |^2 }$} & {$\frac{1}{ 1 + 2 \vec{a} \cdot \vec{x} + | \vec{ a} |^2 |\vec{x} |^2 }$} & $\CI(\vec{ a} + \vec{\iota}\,(\vec{x}))\CI(\vec{x})$ & $|\vec{ a}|^2 \vec{n}_{\vec{a}}(\vec{x}) - {\vec{a}}$ \\
Inversions, $\vec{\iota}$ &{$\frac{\vec{x}}{|\vec{x}|^2 }$} & {$\frac{1}{|\vec{x}|^2}$} & $\CI(\vec{x}) \equiv \id - 2 {\hat x} \!\otimes\! {\hat x}$ & $\vec{\iota}\,(\vec{x})$ \\
\hline
\end{tabular}
\caption{Five classes of conformal diffeomorphisms: Translations, rotations, dilations, special conformal transformations (SCT) and inversions. The first four elements live in the identity component of the group.}
\label{table:conformal_group}
\end{center}
\end{table}

The identity component of the Lorentz group is isomorphic to the identity component of the Euclidean conformal group.\footnote{The isomorphism also extends to the disconnected component of the conformal group and the disconnected component of the Lorentz group involving parity transformations, $\vec{\iota} \mapsto \L_{\vec{\iota}} = \diag(+1,+1,\cdots,+1,-1)$.} For the conformal diffeomorphisms described in Table \ref{table:conformal_group}, the isomorphism is given by
\begin{equation}
\begin{split}\label{CNB-def}
\L_{\vec{\tilde n}_{\vec a}} &= \begin{pmatrix} 1 + \frac{1}{2} |\vec{ a}|^2 & \vec{ a} & \frac{1}{2} |\vec{ a}|^2   \\  \vec{ a}\,^T & {\mathsf 1} & \vec{ a}\,^T \\ - \frac{1}{2} |\vec{ a}|^2 & - \vec{ a} & 1 - \frac{1}{2} |\vec{ a}|^2 \end{pmatrix} , \qquad 
\L_{\vec{m}_R} =  \begin{pmatrix} 1 & \vec{0} & 0 \\ \vec{0}\,^T & R & \vec{0}\,^T \\ 0 & \vec{0} & 1  \end{pmatrix} , \\
\L_{\vec{n}_{\vec a}} &= \begin{pmatrix} 1 + \frac{1}{2} |\vec{ a}|^2 & \vec{ a} & - \frac{1}{2} |\vec{ a}|^2  \\  \vec{ a}\,^T & {\mathsf 1} & - \vec{ a}\,^T \\  \frac{1}{2} |\vec{ a}|^2  & \vec{ a} & 1 - \frac{1}{2} |\vec{ a}|^2  \end{pmatrix} , \qquad\quad
\L_{\vec{a}_\l} =  \begin{pmatrix} \frac{1 + \l^2}{2\l}  & \vec{0} & \frac{1 - \l^2}{2\l}  \\  \vec{0}\,^T & {\mathsf 1} & \vec{0}\,^T \\ \frac{1 - \l^2}{2\l}  & \vec{0} &  \frac{1 + \l^2}{2\l} \end{pmatrix} . 
\end{split}
\end{equation}
For more general elements, the isomorphism is constructed by using the property $\L_{\vec{V}_1} \L_{\vec{V}_2} = \L_{\vec{V}_1 \circ \vec{V}_2}$.

The group isomorphism extends to the corresponding algebras as well. The generators $M_{\mu\nu}$ of the Lorentz algebra satisfy
\begin{equation}
\begin{split}
[ M_{\mu\nu} , M_{\rho\s} ] = 4 i \eta_{[\rho[\mu} M_{\nu]\s]} .
\end{split}
\end{equation}
On the other hand, the generators $T_a$, $K_a$, $D$ and $J_{ab}$ of the Euclidean conformal algebra satisfy
\begin{equation}
\begin{split}\label{conf-algebra}
[ T_a , D  ] &= + i T_a , \qquad [ K_a , D  ] = - i K_a , \qquad [ T_a , K_b ] = - 2 i ( \d_{ab} D + J_{ab} ) , \\
[ J_{ab} , T_c ] &= 2 i \d_{c[a} T_{b]} , \qquad [ J_{ab} , K_c ] = 2 i \d_{c[a} K_{b]} , \qquad [ J_{ab} , J_{cd} ] = 4 i \d_{[c[a} J_{b]d]} , \qquad \text{others} = 0 . 
\end{split}
\end{equation}
It is easy to verify that two algebras are mapped to each other via
\begin{equation}
\begin{split}\label{confgen}
M_{0a} = \frac{1}{2} ( T_a + K_a ) , \qquad M_{d+1,a} = \frac{1}{2} ( K_a - T_a ) , \qquad M_{ab} = J_{ab} , \qquad M_{d+1,0} = D . 
\end{split}
\end{equation}

\subsection{Scattering Amplitudes as Correlators}
\label{sec:amp_correlators}

The group and algebra isomorphism described in the previous section can be used to recast the plane wave operators in a form that highlights their similarity to a conformal primary operator. To see this, we parameterize a null momentum as
\begin{equation}
\begin{split}
\label{mompar}
p^\mu ( \o , \vec{x} ) = \o {\hat q}^\mu(\vec{x})  , \qquad {\hat q}^\mu(\vec{x}) \equiv \left( \frac{1+|\vec{x}|^2}{2} , \vec{x} , \frac{1-|\vec{x}|^2}{2} \right) , \qquad n^\mu \equiv \frac{1}{2} ( 1 , \vec{0} , - 1 ) . 
\end{split}
\end{equation}
Using this parameterization, the operator \eqref{LSZ} can be written as
\begin{equation}
\begin{split}
\label{pwo_def_1}
\CO (\o,\vec{x}) =  - i \ve (\vec{x}) \int_{\mrr^{d+2}} \dt^{d+2} X e^{ - i \o {\hat q} (\vec{x}) \cdot X  } \p^2 \varphi ( X ) . 
\end{split}
\end{equation}

The replacement $p^\mu \to ( \o, \vec{x})$ is, at this stage, simply a notational change. The advantage of doing this can be understood by working out the Lorentz and CPT transformation laws \eqref{Ok_LT} in this notation. We start by noting the property
\begin{equation}
\begin{split}
\label{massless_transformation}
( \L_{\vec{V}} )^\mu{}_\nu p^\nu( \o,\vec{x}) = p^\mu \left( \frac{\o}{\O_{\vec V}(\vec{x})} , \vec{V}(\vec{x}) \right) , \qquad \CR^\mu{}_\nu p^\nu( \o,\vec{x}) = p^\mu( - \o , - \vec{x}) . 
\end{split}
\end{equation}
We next turn to the little group. Lorentz matrices which preserve $r^\mu$ have the form
\begin{equation}
\begin{split}
W_{R,\vec a} = \L_{\vec{m}_R} \L_{\vec{n}_{\vec a}} ~\in~ \ISO(d).
\end{split}
\end{equation}
$\L_{\vec{m}_R}$ generates the $\SO(d)$ rotations and $\L_{\vec{n}_{\vec a}}$ generates the translations within $\ISO(d)$.\footnote{The $d$-dimensional translations within $\ISO(d)$ should not be confused with the $(d\!+\!2)$-dimensional spacetime translations.} Note that $\ISO(d)$ is a non-compact group where the non-compact directions are generated by $\L_{\vec{n}_{\vec a}}$. Since all finite-dimensional unitary irreducible representations must act trivially along the non-compact directions, we have
\begin{equation}
\begin{split}
\label{Dk_prop}
\SD ( W_{R,\vec{a}}) = \SD(R) .
\end{split}
\end{equation}
In other words, $\SD$ is a representation of the maximal compact subgroup $\SO(d)$. 

Next, we discuss $W(\L,p)$ defined in \eqref{W_def}. The Lorentz matrix $L(p)$ defined by \eqref{Lp_def} is given by
\begin{equation}
\begin{split}
\label{Lp_explicit}
L(p) \equiv L(\o,\vec{x}) = \L_{\vec{\tilde{n}}_{\vec{x}}} \L^{-1}_{\vec{a}_{\o}} .
\end{split}
\end{equation}
Using this, we find
\begin{equation}
\begin{split}
W(\L,p) \equiv W(\L_{\vec{V}} ; \o , \vec{x} ) = W_{R_{\vec{V}}(\vec{x}), - \o^{-1} \vec{\chi}_{\vec{V}}(\vec{x})} .
\end{split}
\end{equation}

Finally, we turn to $C$. Using the fact $\CR W_{R,\vec{a}} \CR^{-1} = W_{R,-\vec{a}}$ and \eqref{Dk_prop}, we can rewrite the first equation of \eqref{Ck_prop} as $C \SD(R) C^{-1} = \SD(R)$ which implies that $C$ is proportional to the identity matrix (Schur's lemma). The second equation of \eqref{Ck_prop} then implies that $C = e^{\frac{\pi i}{2} F}$.

Using all of this, we find that the Lorentz and CPT transformation laws for the plane wave operator take the form
\begin{align}
\label{pwo_LT}
U(\L_{\vec{V}}  ) \cdot \CO(\o,\vec{x}) &= \SD \left( R^{-1}_{\vec{V}}(\vec{x}) \right) \CO \left( \frac{\o}{\O_{\vec V}(\vec{x})} , \vec{V}(\vec{x}) \right) , \\
\label{pwo_CPT}
\CPT \cdot \CO (\o,\vec{x}) &= e^{-\frac{\pi i}{2} F}\CO^\dagger (-\o,-\vec{x}) .
\end{align}
The Lorentz transformation property \eqref{pwo_LT} can be equivalently written as
\begin{equation}
\begin{split}
\label{pwo_LT_1}
U(\L_{\vec{V}}  ) \cdot \CO(\o,\vec{x}) = \O_{\vec V}^{-\o\p_{\o}} (\vec{x}) \SD \left( R^{-1}_{\vec{V}}(\vec{x}) \right) \CO  ( \o , \vec{V}(\vec{x})  ) .
\end{split}
\end{equation}
This is almost identical to the transformation law for a conformal primary operator, which is
\begin{equation}
\begin{split}\label{cpo_LT}
U(\L_{\vec{V}}  ) \cdot \SCO(\D,\vec{x}) = \O_{\vec V}^{\D} (\vec{x})\SD \left( R^{-1}_{\vec{V}}(\vec{x}) \right) \SCO ( \D , \vec{V}(\vec{x}) ). 
\end{split}
\end{equation}
Comparing \eqref{cpo_LT} to \eqref{pwo_LT_1}, we see that the plane wave operators represent the scaling dimension as a differential operator $\D=-\o\p_\o$. To see this even more transparently, it is instructive to move to the conformal algebra \eqref{conf-algebra} and study the action of the conformal generators on the plane wave operators. Expanding \eqref{pwo_LT} near the identity, we find 
\begin{align}
\label{Ta_action}
T_a \cdot \CO(\o,\vec{x})  &=  - i \p_a \CO(\o,\vec{x}) , \\
\label{Jab_action}
J_{ab} \cdot \CO(\o,\vec{x}) &= - i  ( - x_a  \p_b + x_b \p_a - i \, \SS_{k\,ab} ) \CO(\o,\vec{x})  , \\
\label{D_action}
D \cdot \CO(\o,\vec{x}) &= - i ( x^a \p_a - \o \p_{\o} ) \CO(\o,\vec{x})  , \\
\label{Ka_action}
K_a  \cdot \CO(\o,\vec{x})  &= - i ( x^2  \p_a - 2 x_a x^b \p_b - 2 x_a ( - \o \p_\o ) - 2 i x^b\, \SS_{k\,ab} ) \CO(\o,\vec{x}) .
\end{align}
These are precisely the transformation laws for a conformal primary operator
\begin{align}
\label{Ta_action_1}
T_a \cdot \SCO(\D,\vec{x})  &=  - i \p_a \SCO(\D,\vec{x}) , \\
\label{Jab_action_1}
J_{ab} \cdot \SCO(\D,\vec{x}) &= - i  ( - x_a  \p_b + x_b \p_a - i \, \SS_{k\,ab} ) \SCO(\D,\vec{x})  , \\
\label{D_action_1}
D \cdot \SCO(\D,\vec{x}) &= - i ( x^a \p_a + \D ) \SCO(\D,\vec{x})  , \\
\label{Ka_action_1}
K_a  \cdot \SCO(\D,\vec{x})  &= - i ( x^2  \p_a - 2 x_a x^b \p_b - 2 x_a \D  - 2 i x^b\, \SS_{k\,ab} ) \SCO(\D,\vec{x}) .
\end{align}
As before, we see that the two are mapped to each other by setting $\D = - \o \p_\o$. 

We end this section by noting that using this new notation, the action of the (spacetime) translation generators $P_\mu$ on the plane wave operators \eqref{translation_action} can be written as
\begin{equation}
\begin{split}
\label{translation_action_1}
P_\mu \cdot \CO(\o,\vec{x}) = - \o {\hat q}_\mu(\vec{x})  \CO(\o,\vec{x}) . 
\end{split}
\end{equation}

\subsection{Conformal Primaries with \texorpdfstring{$\D \in \frac{d}{2}+i\mrr$}{DeltainPS}}
\label{sec:Pasterski_Shao}

Given the remarkable similarity between equations \eqref{Ta_action}--\eqref{Ka_action} and equations \eqref{Ta_action_1}--\eqref{Ka_action_1}, one can hope that there is a simple map between the plane wave operators $\CO$ and the conformal primary operators $\SCO$. This map was determined in \cite{Pasterski:2017kqt}. The basic idea of that work was to construct wavefunctions that transform as conformal primaries, which are then used to construct the operators $\SCO$. For instance, a scalar conformal primary wave function is defined to satisfy
\begin{equation}
\begin{split}
\label{cpwf_eq}
\p^\mu \p_\mu \Uppsi_{\D,\vec{x}}^\pm(X) = 0 , \qquad \Uppsi^\pm_{\D,\vec{x}}  ( \L_{\vec{V}}^{-1} X  ) =  \O_{\vec{V}}^\D (\vec{x}) \Uppsi^\pm_{\D,\vec{V}(\vec{x})}(X) ,
\end{split}
\end{equation}
where the $\pm$ sign corresponds to the positive and negative energy modes, respectively.\footnote{The explicit form of the conformal primary wavefunctions can be determined by solving \eqref{cpwf_eq}, and one finds
$$
\Uppsi^\pm_{\D,\vec{x}}(X) = \frac{ \G(\D) }{  ( \mp i X \cdot {\hat q}(\vec{x}) + \e )^\D }.
$$
}
We use these wave functions to define the conformal primary operators
\begin{equation}
\begin{split}
\SCO^\pm (\D,\vec{x}) \equiv - i \int \dt^{d+2} X \, \Uppsi^\pm_{\D,\vec{x}}(X)  \p^2 \varphi ( X ) .
\end{split}
\end{equation}
The relationship between the operators $\SCO$ and $\CO$ can then be determined using the explicit map between the conformal primary wavefunction and the plane-wave wavefunction, $\Upphi^\pm_p(X) = e^{\pm i p \cdot X}$. The final result of this entire construction is that the plane wave operators and conformal primary operators are related via the Mellin transform
\begin{equation}
\begin{split}
\label{Mellin_Transform}
\SCO^\pm(\D,\vec{x}) = \int_{\mrr_+} \dt \o \o^{\D-1} \CO(\pm \o,\vec{x} ) , 
\end{split}
\end{equation}
and inversely,
\begin{equation}
\begin{split}
\label{Inverse_Mellin_Transform}
\CO(\pm \o,\vec{x}) = \int_{\CC_P} \frac{\dt \D}{2\pi i}  \o^{-\D} \SCO^\pm(\D,\vec{x} ) , \qquad \CC_P \cong \frac{d}{2} + i \mrr . 
\end{split}
\end{equation}
CPT transformations relate the $out$ and $in$ operators to each other
\begin{equation}
\begin{split}
\CPT \cdot \SCO^\pm(\D,\vec{x}) = e^{-\frac{\pi i}{2} F} \SCO^{\mp \dagger}(\D,- \vec{x}) .
\end{split}
\end{equation}
Consistency with CPT symmetry implies that we cannot restrict to a single set of modes. Celestial amplitudes with all $+$ or all $-$ vanish identically (this corresponds to amplitudes with all $out$ particles or all $in$ particles, respectively), and interesting celestial amplitudes involve both operators.

%

\section{Conformal Primaries with General \texorpdfstring{$\D$}{Delta}}
\label{sec:constraints}

This section discusses conformal primary operators with general scaling dimensions $\D \in \mcc$. While we are motivated by the problems outlined in the introduction of the paper, we do not specifically set out to resolve them here. Rather, we will impose some general constraints on the conformal primaries (e.g., completeness and normalizability) and find that this restricts the scaling dimensions of the conformal primary to $\D \in \CC_P$ or $\D \in \mrr$. It will turn out that the conformal primaries with $\D \in \mrr$ automatically resolve the issues above.

We start by generalizing the integration contour used in \eqref{Mellin_Transform} and write\footnote{We do \emph{not} assume that $\CC_\o$ is a continuous deformation of the Mellin contour $\mrr_+$ so there is a priori, no relation between the conformal primary operators defined by \eqref{new_defs} and the ones defined by the Mellin transform.}
\begin{equation}
\begin{split}
\label{new_defs}
\SCO(\D,\vec{x}) = c \int_{\CC_\o} \dt \o \o^{\D-1} \CO(\o,\vec{x} ) .
\end{split}
\end{equation}
We have also included an arbitrary normalization $c$ that we will fix later.

To define the contour integral \eqref{new_defs}, we must analytically continue the plane wave operators to $\o \in \mcc$ to define the contour integrals. This is done by analytically continuing the integral representation \eqref{pwo_def_1} by allowing $\o \in \mcc$ while keeping $X^\mu$ real. We will also assume that the analytic continuation satisfies
\begin{equation}
\begin{split}
\label{prop_1}
\CO(e^{2\pi i} \o , e^{-2\pi i} \vec{x} ) = \CO ( \o , \vec{x} )  \quad \implies \quad \SCO(\D,e^{-2\pi i} \vec{x}) = e^{2\pi i \D} \SCO(\D,\vec{x}) . 
\end{split}
\end{equation}
Importantly, this analytic continuation preserves the Poincar\'e transformation properties \eqref{Ta_action}--\eqref{Ka_action} and \eqref{translation_action_1} of the plane wave operator. The action of the CPT operator \eqref{pwo_CPT} can now be written as
\begin{equation}
\begin{split}
\label{CPT_transform_1}
\CPT \cdot \CO (\o,\vec{x}) &= e^{-\frac{\pi i}{2} F}\CO^\dagger (e^{\mp \pi i} \o , e^{\pm \pi i} \vec{x}) .
\end{split}
\end{equation}
CPT maps $p \to \CR p$, which preserves the product $\o \vec{x}$. Consequently, $\o$ and $\vec{x}$ are rotated by opposite phases above. The choice of sign here is irrelevant due to the property \eqref{prop_1}.

To fix $\CC_\o$, we require that $\SCO(\D,\vec{x})$ forms a complete basis, is normalizable, and has well-defined properties under Lorentz and CPT transformations. The central result of this paper, and the one we will prove in this section, is that these three constraints restrict the conformal primaries to one of two possibilities:
\begin{enumerate}[leftmargin=*]
\label{main_result}
\item[I.] \textbf{Type I Conformal Primaries:} There are two sets of Type I conformal primaries defined by
\begin{equation}
\begin{split}
\label{TypeI_primaries}
\SCO^+ ( \D,\vec{x} ) &= \int_{e^{it_0} \mrr_+} \dt \o \o^{\D-1} \CO ( \o,\vec{x} ) , \qquad \SCO^- ( \D,\vec{x} ) =  \int_{e^{i(t_0-\pi)}\mrr_+} \dt \o \o^{\D-1} \CO ( \o,\vec{x} ) ,
\end{split}
\end{equation}
for any $t_0 \in [0,\pi)$. Note that these contours are unbounded in $\o$. The scaling dimensions of the conformal primaries are restricted to the principle series axis $\D \in \CC_P$. The inverse map is
\begin{equation}
\begin{split}
\label{TypeI_inverse}
\CO ( \o,\vec{x} ) &= \int_{\CC_P} \frac{\dt \D}{2\pi i} \o^{-\D} \SCO^+(\D,\vec{x})  , \qquad \o \in e^{it_0} \mrr_+ , \\
 \CO ( \o,\vec{x} ) &=  \int_{\CC_P} \frac{\dt \D}{2\pi i} \o^{-\D} \SCO^-(\D,\vec{x}) , \qquad \o \in e^{i(t_0-\pi)} \mrr_+  .
\end{split}
\end{equation}

\item[II.] \textbf{Type II Conformal Primaries:} There is one set of Type II conformal primaries defined by
\begin{equation}
\begin{split}
\SCO(\D,\vec{x}) = \int_{\CC_\L} \frac{\dt \o}{2\pi i} \o^{\D-1} \CO( \o , \vec{x}) , \qquad \CC_\L \cong \L e^{i \mrr}  , 
\end{split}
\end{equation}
for any $\L \in \mrr_+$. Note that this contour is bounded in $\o$. The scaling dimensions of the conformal primaries are restricted to $\D \in \mrr$, and the inverse transformation is given by
\begin{equation}
\begin{split}
\CO(\o,\vec{x}) = \int_{\mrr} \dt \D\, \o^{-\D} \SCO(\D,\vec{x} ).
\end{split}
\end{equation}
\end{enumerate}
The Pasterski-Shao conformal primaries are of Type I with $t_0=0$. Type I conformal primaries with $t_0 \neq 0$ can be analytically continued to those with $t_0=0$, so the parameter $t_0$ is, in this sense, topological. The same applies to the parameter $\L$ that defines the Type II conformal primaries.

In the rest of this section, we prove this result.

\subsection{Completeness}
\label{sec:completeness}

The first requirement we study is the completeness of the conformal primary operators (as a basis on the space of asymptotic states). We do this by requiring that the map \eqref{new_defs} is invertible. Then, since the plane wave operators form a complete basis, so do the conformal primary operators.

The inverse basis transform is obtained by expanding the plane wave operators in a basis of eigenfunctions $\o^{-\D}$ of the derivative operator $-\o\p_\o$ which has eigenvalue $\D$. The expansion takes the general form
\begin{equation}
\begin{split}\label{new_defs_inverse}
\CO(\o,\vec{x}) = c' \int_{\CC_\D} \dt \D \o^{-\D} \SCO(\D,\vec{x}) , 
\end{split}
\end{equation}
where the integral is over some contour $\CC_\D$ in complex $\D$ space and $c'$ is some yet-to-be fixed normalization constant. Substituting the inverse formula \eqref{new_defs_inverse} into \eqref{new_defs} and vice versa, we find the conditions
\begin{equation}
\begin{split}
\label{req_1}
c c' \int_{\CC_\D} \dt \D \o^{-\D} \o'^{\D-1} = \d_{\CC_\o} ( \o' , \o )  , \qquad c c' \int_{\CC_\o} \dt \o \o^{\D-\D'-1}  = \d_{\CC_\D} ( \D' , \D )  ,
\end{split}
\end{equation}
where $\d_{\CC}$ is the Dirac delta function along the contour $\CC$, normalized according to
\begin{equation}
\begin{split}
\label{generalized_delta_func}
\int_\CC \dt z' \d_{\CC}(z',z) f(z') = f(z) . 
\end{split}
\end{equation}
To process \eqref{req_1}, working with a new variable
\begin{equation}
\begin{split}
\label{tdef}
t = \ln \o
\end{split}
\end{equation}
is convenient. In terms of the $t$, \eqref{req_1} reads
\begin{equation}
\begin{split}
\label{req_3}
c c' \int_{\CC_t} \dt t e^{(t-t')(\D-\D')}  = \d_{\CC_\D} ( \D'  , \D )  , \qquad c c'  \int_{\CC_\D} \dt \D\,e^{-(\D-\D')(t-t')}  = \d_{\CC_t} ( t' , t ) . 
\end{split}
\end{equation}
Comparing this to the standard integral representation of the standard Dirac delta function,
\begin{equation}
\begin{split}\label{DDF_integral_rep}
\frac{1}{2\pi i} \int_{i \mrr} \dt x e^{(x-x')(k-k')} = \d_\mrr(k,k') = \d ( k - k' ) .
\end{split}
\end{equation}
we see that \eqref{req_3} holds if the function $( t - t' ) ( \D - \D ') \in i\mrr$ for all $t,t'\in\CC_t$ and $\D,\D'\in\CC_\D$.\footnote{More precisely, we need the function $f_{\D,\D',t'}(t) = ( t - t' ) ( \D - \D ')$ to be an invertible map from $\CC_t \to i \mrr$ for all $\D,\D'
 \in \CC_\D$ and $t' \in \CC_t$ and similarly, we need the function $g_{t,t',\D'}(\D) = ( t - t' ) ( \D - \D ')$ to be an invertible map from $\CC_\D \to i \mrr$ for all $t,t' \in \CC_t$ and $\D' \in \CC_\D$.} This is an incredibly strong constraint that almost completely fixes both contours. For example, keeping $t'$, $\D$, $\D'$ fixed with $\D \neq \D'$, we find
\begin{equation}
\begin{split}
t = t'  + i (\D-\D')^{-1} \mrr 
\end{split}
\end{equation}
which immediately implies that $\CC_t$ is a straight-line contour. A similar computation shows that $\CC_\D$ is also a straight line and that the contours $\CC_t$ and $\CC_\D$ are perpendicular to each other. Finally, comparing the normalizations of \eqref{req_3} to that in \eqref{DDF_integral_rep} we find $cc'=\frac{1}{2\pi i}$. In summary, we have
\begin{equation}
\begin{split}
\CC_\D \cong \D_0 + e^{i\a} \mrr , \qquad \CC_t \cong t_0 + i e^{-i\a} \mrr , \qquad cc'=\frac{1}{2\pi i} . 
\end{split}
\end{equation}
for some fixed $\D_0,t_0 \in \mrr$ and $\a \in [0,\pi)$. There are two qualitatively distinct types of contours depending on whether or not $\a$ vanishes.
\begin{itemize}[leftmargin=*]
\item \textbf{Type I Contours ($\a \neq 0$):} In this case, without loss of generality, we can choose $\D_0 \in \mrr$ and replace $t_0 \to i t_0 $ where $ t_0 \in \mrr$. The contours are
\begin{equation}
\begin{split}
\label{TypeI}
\text{Type I:} \qquad \CC_\D \cong \D_0 + e^{i\a} \mrr , \qquad \CC_t \cong i t_0 + i e^{-i\a} \mrr  \quad \stackrel{\eqref{tdef}}{\implies} \quad \CC_\o \cong e^{i t_0} e^{i e^{-i\a} \mrr}   .
\end{split}
\end{equation}
The Type I contours in $\D$ and $\o$ are shown in Figure \ref{fig:TypeI} for three separate cases: $\a\in(0,\frac{\pi}{2})$, $\a=\frac{\pi}{2}$, and $\a \in (\frac{\pi}{2},\pi)$. In each of these cases, the contour $\CC_\o$ is unbounded. 
\begin{figure}[ht!]
\begin{center}
\begin{tabular}{ccc}
\includegraphics[width=0.3\linewidth]{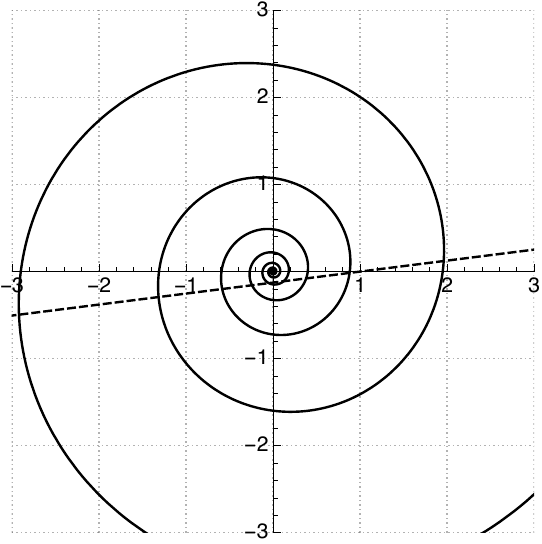} &
\includegraphics[width=0.3\linewidth]{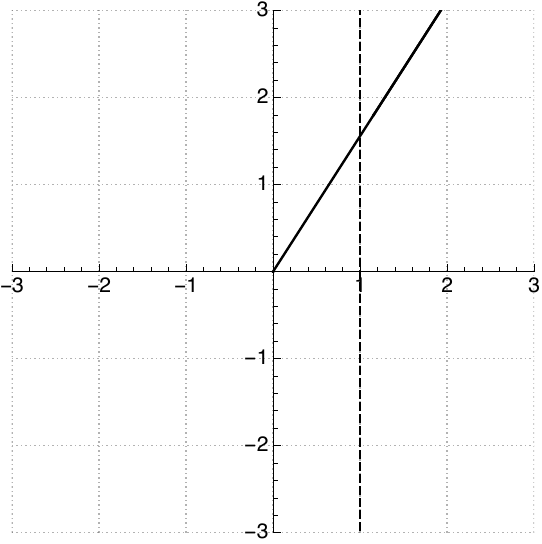} &
\includegraphics[width=0.3\linewidth]{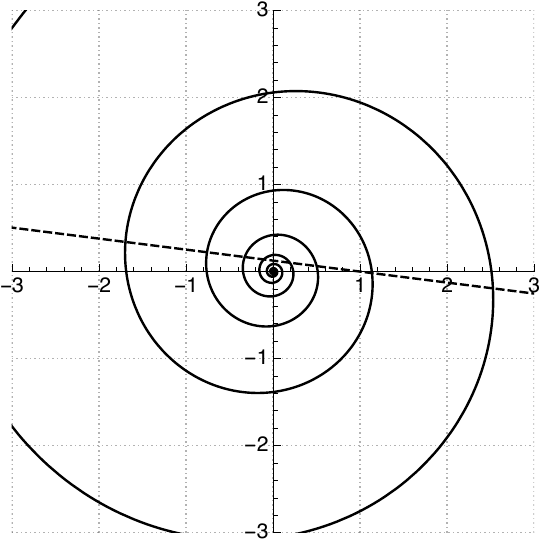} \\
$\a \in (0,\frac{\pi}{2})$ & $\a=\frac{\pi}{2}$ & $\a \in ( \frac{\pi}{2} , \pi )$
\end{tabular}
\caption{Type I contours in $\o$ (solid) and $\D$ (dashed) for $\a \in (0,\pi)$ and $\D_0=t_0=1$.}
\label{fig:TypeI}
\end{center}
\end{figure}

For the Type I contours, we choose the normalization constants to be
\begin{equation}
\begin{split}
c=1, \qquad c' = \frac{1}{2\pi i} .
\end{split}
\end{equation}

\item \textbf{Type II Contours ($\a = 0$):} In this case, without loss of generality, we can choose $t_0 \in \mrr$ and replace $\D_0 \to i \D_0$ where $\D_0\in\mrr$. The contours, in this case, are
\begin{equation}
\begin{split}
\label{TypeII}
\text{Type II:} \qquad \CC_\D \cong \mrr + i \D_0 , \qquad \CC_t \cong t_0 + i \mrr  \quad \stackrel{\eqref{tdef}}{\implies} \quad \CC_\o \cong \L e^{i\mrr} ,
\end{split}
\end{equation}
where $\L = e^{t_0} \in \mrr_+$. The Type II contours in $\D$ and $\o$ are shown in Figure \ref{fig:TypeII}. The contour $\CC_\o$ in this case is bounded.
\begin{figure}[ht!]
\begin{center}
\includegraphics[width=0.3\linewidth]{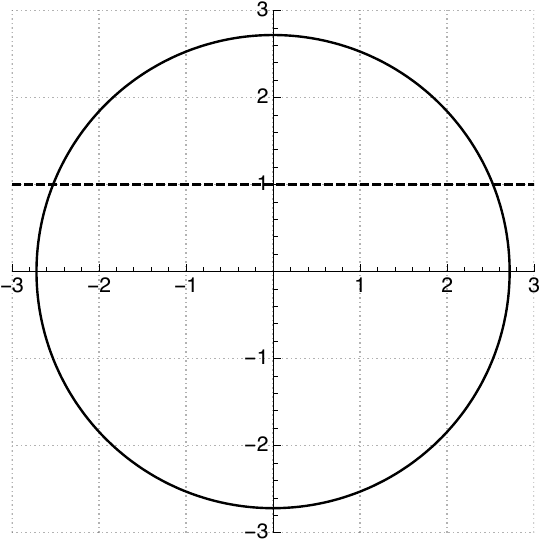}
\caption{Type II contours in $\o$ (solid) and $\D$ (dashed) for $\a=0$ and $\D_0=t_0=1$.}
\label{fig:TypeII}
\end{center}
\end{figure}
For the Type II contours, we choose the normalization constants to be 
\begin{equation}
\begin{split}
c = \frac{1}{2\pi i} , \qquad c' = 1 . 
\end{split}
\end{equation}

\end{itemize}

\subsection{Consistency with Symmetries}

The next requirement we study is consistency with Lorentz and CPT symmetries and with soft theorems (asymptotic symmetries).

\subsubsection{Lorentz Transformations} We start by requiring that $\SCO$ transform as a conformal primary. Consider the action of the generator $D$ on the $\SCO$.\footnote{Ensuring that $D$ acts correctly automatically implies that the remaining conformal generators act correctly.} Using \eqref{D_action}, we find
\begin{equation}
\begin{split}
\label{SCO_D_action}
D \cdot  \SCO(\D,\vec{x})  &= - i c \int_{\CC_\o} \dt \o \o^{\D-1} ( x^a \p_a - \o \p_{\o} ) \CO (\o,\vec{x})  \\ 
&= - i ( x^a \p_a + \D )  \SCO(\D,\vec{x})  + i c \int_{\CC_\o} \dt \o \p_\o [ \o^\D \CO(\o,\vec{x}) ] ,
\end{split}
\end{equation}
where we integrated by parts in $\o$ in the second equality. The first term is what we want for a conformal primary (see \eqref{D_action_1}), so we must impose the boundary condition
\begin{equation}
\begin{split}
\label{boundary_contour_restriction}
\o^\D \CO(\o,\vec{x}) |_{\p \CC_\o} = 0 . 
\end{split}
\end{equation}
We analyze this boundary condition for the Type I and II contours separately.
\begin{itemize}[leftmargin=*]
\item \textbf{Type I:} These contours start at $\o=0$ and end at $\o=\infty$, so \eqref{boundary_contour_restriction} is equivalent to
\begin{equation}
\begin{split}
\label{TypeI_restriction}
\lim_{\o \to 0} \o^\D \CO(\o,\vec{x}) = 0 , \qquad \lim_{\o \to \infty} \o^\D \CO(\o,\vec{x}) . 
\end{split}
\end{equation}
The IR constraint (at $\o=0$) can be immediately checked using soft theorems, which imply that the plane wave operators satisfy
\begin{equation}
\begin{split}
\CO(\o,\vec{x}) = O(\o^{-1}) \quad \text{as} \quad \o \to 0.
\end{split}
\end{equation}
The IR limit of \eqref{TypeI_restriction} is therefore satisfied if and only if\footnote{The equality is included since, in this case, $\o^\D \CO(\o,\vec{x})$ oscillates rapidly as $\o \to 0$, which vanish when integrated against suitably normalized wavepackets (Riemann-Lebesgue lemma).}
\begin{equation}
\begin{split}\label{delta_bound}
\text{Re}\,\D \geq 1 . 
\end{split}
\end{equation}
However, \eqref{TypeI} implies that $\text{Re} \,\CC_\D \cong \D_0 + \cos \a\, \mrr$ so the bound \eqref{delta_bound} holds if and only if $\a = \frac{\pi}{2}$ and $\D_0 \geq 1$.

The UV constraint in \eqref{TypeI_restriction} is non-trivial and can only be checked on a case-by-case basis. It implies, for instance, that the plane wave operators (and hence the scattering amplitudes) must be UV complete and that they must be sufficiently soft in the UV. Interestingly, string amplitudes seem to satisfy this constraint \cite{Donnay:2023kvm}. We do not comment on this further.

\item \textbf{Type II:} In this case, \eqref{boundary_contour_restriction} reads
\begin{equation}
\begin{split}
\lim_{\eta \to \pm\infty} e^{(t_0 + i \eta) (\nu + i \D_0)} \CO(e^{ t_0 } e^{ i \eta } ,\vec{x})  = 0 .
\end{split}
\end{equation}
The plane wave operator $\CO$ remains finite in this limit (since $|\o|$ is fixed and finite on this contour). Hence, this holds if and only if $\D_0 = 0$ (as before, if $\D_0 = 0$, the function oscillates rapidly as $\eta \to \pm \infty$ and therefore vanishes when integrated against suitably normalized wavepackets).

\end{itemize}

\subsubsection{CPT} We next turn to consistency with CPT transformations. Using \eqref{CPT_transform_1}, we find
\begin{equation}
\begin{split}
\CPT \cdot \SCO (\D,\vec{x}) &= e^{-\frac{\pi i}{2} F  } \left( c  \int_{\CC_\o} \dt \o \o^{\D-1} \CO (e^{- \pi i} \o , e^{\pi i} \vec{x})  \right)^\dagger . 
\end{split}
\end{equation}
Without loss of generality, we have chosen to work with the top sign in \eqref{CPT_transform_1}. 
To return this to the form \eqref{new_defs}, we need to change the integration variable $\o \to e^{\pi i} \o$. Depending on the type of contour, we have two cases:
\begin{itemize}[leftmargin=*]
\item \textbf{Type I:} In this case, $\CC_\o$ is not invariant under $\o \to e^{\pi i} \o$. Rather, it rotates to a new Type I contour $\CC'_\o$ defined by the parameter $t_0' = t_0 - \pi$. We use this rotated contour to define a new conformal primary operator
\begin{equation}
\begin{split}
\label{wt_Ok_def}
\SCO' (\D,\vec{x}) \equiv \int_{\CC'_\o} \dt \o \o^{\D-1} \CO ( \o , \vec{x}) .
\end{split}
\end{equation}
Since we have two sets of conformal primaries, we change our notation and replace $\SCO \to \SCO^+$ and $\SCO' \to \SCO^-$. Under CPT transformations,
\begin{equation}
\begin{split}
\CPT \cdot \SCO^\pm (\D,\vec{x}) &= e^{-\frac{\pi i}{2} F  } e^{- \pi i \D^\star } \SCO^{\mp\dagger} (\D, e^{\pi i}  \vec{x}) .
\end{split}
\end{equation}

\item \textbf{Type II:} In this case, $\CC_\o$ is invariant under $\o \to e^{\pi i} \o$, so there is no need to introduce a second set of operators. Under CPT transformations,
\begin{equation}
\begin{split}
\CPT \cdot \SCO(\D,\vec{x}) = e^{-\frac{\pi i}{2} F} e^{ \pi i \D^\star } \SCO^\dagger(\D, e^{\pi i} \vec{x})  . 
\end{split}
\end{equation}

\end{itemize}

\subsection{Normalizability}
\label{sec:norm}

The final requirement we study is that of normalizability. We impose this by requiring that the two-point function of the operators is finite. 
\begin{itemize}[leftmargin=*]
\item \textbf{Type I:} We have four Type I primaries whose two-point function we can consider: $\SCO^\pm$ and ${\bar \SCO}^\pm$. These are the $out$ and $in$ operators and their CPT conjugates. The only non-vanishing two-point functions are of the form $\avg{ \SCO^+ {\bar \SCO}^- }$ or $\avg{ \SCO^- {\bar \SCO}^+ }$. CPT transformations relate the two, so we only need to check the normalizability for one of them. Consider
\begin{equation}
\begin{split}\label{TypeI_two_pt}
\avg{ \SCO^+(\D,\vec{x}) {\bar \SCO}^-(\D',\vec{x}\,' ) }  &= \int_{\CC_\o} \dt \o \o^{\D-1}  \int_{\CC'_\o} \dt \o' \o'^{\D'-1} \avg{ \CO(\o,\vec{x}) {\bar \CO}(\o',\vec{x}\,') } .
\end{split}
\end{equation}
The integrand is the usual $1 \to 1$ scattering amplitude, given by the identity part of the $S$-matrix. For real $\o$, $\o'$ this is given by\footnote{The RHS of \eqref{1to1_amp} is $ (2\pi)^{d+1} (2|{\bf p}|) \d^{(d+1)} ( {\bf p} - {\bf p}' )$ written out in the $(\o,\vec{x})$ parameterization.}
\begin{equation}
\begin{split}\label{1to1_amp}
\avg{ \CO(\o,\vec{x}) {\bar \CO}( \o',\vec{x}\,') }  &= 2  (2\pi)^{d+1} ( - \o \o' )^{\frac{1}{2}(1-d)} \d ( \o + \o' ) \d^{(d)} ( \vec{x} - \vec{x}\,' )   .
\end{split}
\end{equation}
We analytically continue this to complex $\o$, $\o'$ by replacing $\d ( \o + \o' ) \to \d_{\CC_\o} ( \o , e^{\pi i} \o' )$. Using this, we find
\begin{equation}
\begin{split}
\avg{ \SCO^+(\D,\vec{x}) {\bar \SCO}^-(\D',\vec{x}\,' ) } &= 2 e^{\pi i(\D-d )}    (2\pi)^{d+1}  \d^{(d)} ( \vec{x} - \vec{x}\,' )     \int_{\CC'_\o} \dt \o'\o'^{\D+\D-d-1}  .
\end{split}
\end{equation}
The integral along the Type I contour is 
\begin{equation}
\begin{split}
 \int_{\CC'_\o} \dt \o' \o'^{\D+\D'-d-1}  = \int_{-\infty}^\infty \dt \eta e^{( \eta + i ( t_0 - \pi )  ) (2\D_0-d +i (  \nu + \nu'  ) )}  . 
\end{split}
\end{equation}
This integral is convergent (in the distributional sense) if and only if $\D_0 = \frac{d}{2}$ and in this case, we find
\begin{equation}
\begin{split}
 \int_{\CC'_\o} \dt \o' \o'^{\D+\D'-d-1}  = 2\pi \d ( i ( \D + \D' - d ) ) .
\end{split}
\end{equation}
It follows that normalizable Type I conformal primaries live on the principle series axis, $\D \in \CC_P$. The two-point function is given by
\begin{equation}
\begin{split}
\avg{ \SCO^+(\D,\vec{x}) {\bar \SCO}^-(\D',\vec{x}\,' ) } &= 2 e^{\pi i(\D-d )}    (2\pi)^{d+2}  \d ( i ( \D + \D' - d ) )  \d^{(d)} ( \vec{x} - \vec{x}\,' ) .
\end{split}
\end{equation}

\item \textbf{Type II:} We have two Type II conformal primaries that we can consider: $\SCO$ and ${\bar \SCO}$. The only non-vanishing two-point function in this case is 
\begin{equation}
\begin{split}\label{TypeII_two_pt}
\avg{ \SCO(\D,\vec{x}) {\bar \SCO}(\D',\vec{x}\,' ) } = \int_{\CC_\o} \frac{\dt \o}{2\pi i}  \int_{\CC_\o} \frac{\dt \o'}{2\pi i}  \o^{\D-1}  \o'^{\D'-1} \avg{ \CO(\o,\vec{x}) {\bar \CO}(\o',\vec{x}\,') }  . 
\end{split}
\end{equation}
We now analytically continue the Dirac delta function in \eqref{1to1_amp} symmetrically by replacing
\begin{equation}
\begin{split}
\d( \o + \o' ) ~\to~ \frac{1}{2} [ \d_{\CC_\o} ( \o , e^{\pi i} \o' ) + \d_{\CC_\o} ( \o' , e^{\pi i} \o ) ]  .
\end{split}
\end{equation}
Using this, we find
\begin{equation}
\begin{split}
\avg{ \SCO(\D,\vec{x}) {\bar \SCO}(\D',\vec{x}\,' ) } &=  [ e^{\pi i \D} + e^{\pi i \D'} ]   (2\pi)^{d-1} \d^{(d)} ( \vec{x} - \vec{x}\,' )  \int_{\CC_\o} \dt \o  \o^{\D+\D'-d-1} .
\end{split}
\end{equation}
The integral along the Type II contour is of the form
\begin{equation}
\begin{split}
\int_{\CC_\o} \dt \o  \o^{\D+\D'-d-1}  = i \int_{-\infty}^\infty \dt \eta  e^{(t_0+i\eta)(\D+\D'-d)}  = 2\pi i \d ( \D + \D' - d ) . 
\end{split}
\end{equation}
This is already finite (in the distributional sense), so we find no new constraint. Their two-point function is given by
\begin{equation}
\begin{split}
\avg{ \SCO(\D,\vec{x}) {\bar \SCO}(\D',\vec{x}\,' ) } &=   i [ e^{\pi i \D} + e^{\pi i \D'} ]   (2\pi)^d \d ( \D + \D' - d ) \d^{(d)} ( \vec{x} - \vec{x}\,' )  . 
\end{split}
\end{equation}
\end{itemize}
Collecting everything together, we have proved the result described at the top of this section.

\section{Comments}
\label{sec:summary}

In the previous section, we demonstrated that in addition to the conformal primary operators constructed by Pasterski and Shao \cite{Pasterski:2017kqt} which have $\D \in \CC_P$, there exists another complete, normalizable basis of conformal primary operators that are defined as
\begin{equation}
\begin{split}
\label{map2_summary}
\SCO(\D,\vec{x}) = \int_{\L e^{i \mrr}} \frac{\dt \o}{2\pi i} \o^{\D-1} \CO( \o , \vec{x})  , \qquad \CO(\o,\vec{x}) = \int_{\mrr} \dt \D\, \o^{-\D} \SCO(\D,\vec{x} ) .
\end{split}
\end{equation}
Note that the scaling dimensions are restricted to $\D \in \mrr$.

So far, we have not made any assumptions regarding the analytic structure of the plane wave operators. We can now use the fact that in a neighborhood of the $\o=0$, the creation and annihilation operators admit a Laurent expansion of the form
\begin{equation}
\begin{split}
\label{Laurent_Exp}
\CO( \o , \vec{x}) = \sum_{n\in\mzz_{\geq0}} \o^{n-2} \CO^{(n)}(\vec{x}).
\end{split}
\end{equation}
The $\o^{-1}$ pole at $\o=0$ are dictated by soft theorems. We have also included the double pole $\o^{-2}$, which exists only for the gravitational plane wave operators \cite{Guevara:2021abz}. Taking $\L$ to be sufficiently small so that the Laurent expansion \eqref{Laurent_Exp} is valid along the contour $\CC_\L$, we find that the conformal primaries \eqref{map2_summary} have the form
\begin{equation}
\begin{split}
\SCO(\D,\vec{x}) = \sum_{n\in\mzz_{\geq0}} \CO^{(n)} (\vec{x}) \d(\D+n-2) . 
\end{split}
\end{equation}
The conformal primaries are delta-function localized to $\D = 2 - \mzz_{\geq0}$ and the coefficients can be extracted by defining the following operators for $k \in \mzz$,
\begin{equation}
\begin{split}
\wt{\SCO}(k,\vec{x}) \equiv \lim_{\ve \to 0^+} \int_{k-\ve}^{k+\ve} \dt \D \SCO(\D,\vec{x})  = \begin{cases}
\CO^{(2-k)} (\vec{x})  , & k \leq 2 , \\
0 , & k > 2 .
\end{cases}
\end{split}
\end{equation}
Alternatively, we can directly construct the operators $\wt{\SCO}$ by integrating the plane wave operators over the compact contour $\L e^{i[0,2\pi]}$ instead of the one used in \eqref{map2_summary},
\begin{equation}
\begin{split}\label{Otilde_def}
\wt{\SCO}(k,\vec{x}) = \int_{\L e^{i[0,2\pi]}} \frac{\dt \o}{2\pi i} \o^{k-1} \CO( \o , \vec{x})  . 
\end{split}
\end{equation}
It follows from this discussion that even though generically the conformal primaries defined by \eqref{map2_summary} have $\D \in \mrr$, in the case when the plane wave operators admit a Laurent expansion of the form \eqref{Laurent_Exp}, we can instead construct a complete basis of conformal primary operators with $\D \in 2 - \mzz_{\geq 0}$.
The operators constructed by \eqref{Otilde_def} are precisely the conformally soft operators discussed in \cite{Donnay:2018neh, Pate:2019mfs, Puhm:2019zbl} and are related to the $\msw_{1+\infty}$ generators constructed in \cite{Strominger:2021lvk}. For example, the operators $H^k(z,\bz)$ and $O^{a+}_k(z,\bz)$ appearing in \cite{Strominger:2021lvk} ($z$ is the stereographic complex coordinate on the celestial 2-sphere in four dimensions) are precisely the operators \eqref{Otilde_def} constructed from a positive-helicity graviton and positive-helicity gluon plane wave operator, respectively.

We leave a deeper exploration of the conformal primaries \eqref{map2_summary} for future work. One interesting aspect of the operators \eqref{map2_summary} that merits exploration is their operator product expansion. To the extent that a sensible OPE exists, one expects it to take the form
\begin{equation}
\begin{split}\label{2OPE}
\SCO(\D_1,\vec{x}_1)  \SCO(\D_2,\vec{x}_2) \quad \xrightarrow{ \vec{x}_1 \to \vec{x}_2 } \quad \sum_{\D_3} \frac{C(\D_1,\D_2,\D_3)}{|\vec{x}_{12}|^{\D_1+\D_2-\D_3}}  [  \SCO(\D_3,\vec{x}_2) + \text{descendents} ].
\end{split}
\end{equation}
This OPE structure for the conformal primaries results from collinear factorization of scattering amplitudes. For the conformal primaries with $\D \in \CC_P$, the OPE has been explored in \cite{He:2015zea, Pate:2019lpp, Adamo:2021zpw, Himwich:2021dau}.\footnote{The full operator product structure of these conformal primaries is far more complicated and is an area of active research. Scattering amplitudes admit multi-collinear factorization leading to non-trivial OPEs involving more than two operators that cannot be obtained directly from \eqref{2OPE} \cite{Ebert:2020nqf, Ball:2023sdz}. This is rather different from the usual OPE structure of a CFT.} One open problem regarding this OPE is that it is not, in general, unambiguously defined. For example, \cite{Pate:2019lpp} studied the OPE in the `holomorphic limit' $z_{12} \to 0$ keeping $\bz_{12}$ fixed, which is well-defined. The opposite anti-holomorphic limit is also well-defined, but any limit in which $z_{12}$ and $\bz_{12}$ vanish simultaneously is not! This issue can be traced back to the fact that the soft limits of a scattering amplitude involving multiple gluons and gravitons depend on the order in which the gluons are taken to be soft. This ambiguity in the order of soft limits was ``fixed'' in the works mentioned above by simply choosing a particular ordering (positive helicity gluons and gravitons are taken to be soft \emph{before} their negative helicity counterparts). This choice seems arbitrary, and one would like to find a better way to deal with this issue. Interestingly, the conformal primaries we propose \eqref{map2_summary} resolve this. These operators are defined by integrating over a circular contour with radius $\L$. Multiple operator insertions will involve contours defined by radii $\L_i$, which can keep track of the order of softness. More precisely, the operator with the smallest value of $\L_i$ is considered soft \emph{first}. We plan to pursue this further in future work.

\section*{Acknowledgements}

We thank Clifford Cheung, Temple He, Alok Laddha, Daniel Kapec, Andrea Puhm, and Ana-Maria Raclariu for useful discussions and Temple He for comments on the manuscript. P.M. is supported by the European Research Council (ERC) under the European Union’s Horizon 2020 research and innovation programme (grant agreement No 852386). During the initial stages of this work, P. M. was supported by STFC consolidated grants ST/T000694/1 and ST/P000681/1.

\bibliographystyle{utphys}
\bibliography{ccft-integer-bib}

\end{document}